\documentstyle[prb,aps,twocolumn]{revtex}

\input epsfig.sty
\begin{document}

\advance\textheight by 0.2in
\twocolumn[\hsize\textwidth\columnwidth\hsize\csname@twocolumnfalse%
\endcsname

\draft
\begin{flushright}
{\tt to appear in Phys. Rev. B }
\end{flushright}

\title{Nonlinear response scaling of the two-dimensional XY spin glass in the
Coulomb-gas representation}
\author{Enzo Granato}
\address{Laborat\'orio Associado de Sensores e Materiais, \\
Instituto Nacional de Pesquisas Espaciais,\\
12201 S\~{a}o Jos\'{e} dos Campos, SP, Brazil}
\maketitle

\begin{abstract}
Vortex critical dynamics of the two dimensional XY spin glass is studied by
Monte Carlo methods in the Coulomb-gas representation. A scaling analysis of
the nonlinear response is used to calculate the correlation length exponent 
$\nu $ of the zero-temperature glass transition. The estimate, $\nu =1.3(2)$, 
is in agreement with a recent estimate in the phase representation using
the same analysis and indicates that the relevant length scale for vortex
motion is set by the spin-glass correlation length and that spin and
chiralities may order with different correlation length exponents.
\end{abstract}


]

It is well known that vector spin glasses, such as the XY spin glass, have a
chirality order parameter with Ising-like symmetry in addition to the
continuous degeneracy associated with global spin rotation \cite{villain}.
Chirality arises from quenched in vortices due to frustration effects in
each elementary cell of the lattice which contains and odd number of
antiferromagnetic bonds. The interplay between spin and chiral variables has
always received considerable attention because of the possibility of
separate spin-glass and chiral-glass ordering due to freezing of spins and
chiral variables, respectively \cite{kawamura,ray,by96,jain,hilhorst}. 
Possible separation of spin and chiral variables also arise in the frustrated XY 
model with weak disorder \cite{gk86,benakli}. 
While in three dimensions  the existence of a finite temperature transition is under
current investigation \cite{kawali,wengel,grempel}, in two dimensions there is a
consensus that the transition only occurs at zero temperature. Associated
with the zero temperature transition there is a correlation length which
increases with decreasing temperatures as $\xi \propto T^{-\nu }$. However, the
possibility of different spin and chiral glass short range correlation
lengths $\xi _{s}$ and $\xi _{c}$ with different critical exponents $\nu
_{s} $ and $\nu _{c}$ has not been resolved satisfactorily.

First evidence that spin and chiral glass correlation length exponents are
different in two dimensions were reported by Kawamura and Tanemura \cite
{kawamura} from domain wall calculations. Various estimates of these
exponents give approximate values \cite{kawamura,ray,by96,jain} of $\nu
_{s}\sim 1$ and $\nu _{c}\sim 2$ but the errorbars are usually quite large
and a single exponent scenario may not be ruled out which would be
consistent with an analytical work for a particular type of disorder distribution
\cite{hilhorst} and more recent numerical
work \cite{kosterlitz} on domain wall scaling behavior at zero temperature.
These calculations are usually performed in a representation of the XY spin
glass model in terms of the orientational angle of the two-component XY
spins. In this representation, spin glass order can be directly identified
as a long range order in appropriate phase correlations while the chiral
variables are built from nearest neighbor phase correlations. In numerical
simulations, the dynamics of the chiral variables are then determined by the
phase variables and equilibration problems may prevent an adequate study of
the vortex correlations in the system.

An alternative approach which allows to study the vortex dynamics directly
can be obtained from the Coulomb-gas representation \cite{hyman}. Recently,
Bokil and Young \cite{by96} used Monte Carlo simulations in the vortex
representation to obtain an estimate of the chiral glass correlation length
exponent and found $\nu _{c}=1.8\pm 0.3$. This agrees with previous
estimates within the errorbar. It also supports the scenario in which spin
and chiral glass variables order with different critical exponents if one
accepts earlier estimates of the spin glass exponent $\nu _{c}\sim 1$,
obtained in the phase representation. However, it should also be of interest
a determination of the spin glass correlation length exponent from
simulations in the vortex representation as it is not completely clear how
such length scale shows up in the dynamical behavior of vortices. In
particular, since the XY spin glass model is currently been used as a model
for granular high-$T_{c}$ superconductors containing '$\pi $ ' junctions 
\cite{kawali,kusmartsev,dominguez}, which leads to quenched in vortices even
in the absence of external magnetic field, a natural question arises as to
which correlation length, $\xi _{s}$ or $\xi _{c}$, is actually probed by
transport measurements. In the measurements, the response of the vortices to
an applied force can be observed as the voltage response to an applied
driving current which acts as a Lorentz force on the vortices 
\cite{hyman,fisher91}. The vortex response, or resistive behavior, is therefore 
determined by vortex mobility and the current-voltage scaling is expected to
be controlled by the relevant divergent length scale \cite{hyman} which could be 
either $\xi _{s}$ or $\xi _{c}$. 

The  question of the relevant correlation length for vortex response is also 
of interest for the  three dimensional XY spin glass. Recent simulations of the 
vortex dynamics \cite{wengel} in three dimensions showed evidence of a resistive 
phase transition at finite temperatures which was attributed to glass ordering of
chiralities while the spins remain disordered. This would be  consistent with the
scenario of a finite temperature chiral glass transition in the absence of 
spin glass transition which has been proposed previously \cite{kawali} 
from calculations in the phase representation of the XY spin glass model. This 
interpretation however is only justified if the relevant length scale for vortex 
dynamics is determined by the chiral glass correlation length $\xi _{c}$. On the 
other hand, since it is well known that vortex motion leads to phase  incoherence, 
one expects that vortex dynamics should probe the spin glass  correlation length 
$\xi _{s}$  and therefore the resistive transition should correspond instead to a 
spin glass transition at finite temperatures.  In fact, this is supported by   a 
more recent domain-wall calculations suggesting a spin glass transition at finite 
temperatures in three dimensions \cite{grempel}. The present study of the vortex
response in two dimensions may help to clarify this point as it is well known that in 
this case both spin and chiral glass ordering only occurs at zero temperature and
so the scaling analysis involve less unknown parameters. 

In the absence of a precise agreement among the various studies of the XY spin glass
model, as mentioned above, and in view of its relevance for vortex dynamics, the
additional numerical results presented below may help to settle some issues. 

In this work, we study the vortex critical dynamics of the two dimensional
XY spin glass by Monte Carlo methods in the Coulomb-gas representation \cite{hyman}. A
scaling analysis of the nonlinear response is used to calculate the
correlation length exponent $\nu $ of the zero-temperature glass transition.
The estimate, $\nu =1.3(2)$, is in agreement with a recent estimate in the
phase representation \cite{eg98} using the same analysis and indicates that
the relevant length scale for vortex motion is set by the spin-glass
correlation length $\xi _{s}$ and that spin and chiralities may order with
different correlation length exponents.

We consider the XY spin glass on a square lattice defined by the Hamiltonian

\begin{equation}
H=\sum_{<ij>}J_{ij}\ s_{i}\cdot s_{j}=-J\sum_{<ij>}\cos (\theta _{i}-\theta
_{j}-A_{ij})
\end{equation}
where $\theta _{i}$ is the phase of a two-component classical spin of unit
length, $s_{i}=(\cos \theta _{i},\sin \theta _{i})$, $J>0$ is a coupling
constant and $\ A_{ij}$ has a binary distribution, $0$ or $\pi $ , with
equal probability, corresponding to a coupling constant $J_{ij}=-J$ or $J$,
respectively, between $s_{i}$ and $s_{j}$ spins. The sum is over all
nearest-neighbor pairs. This Hamiltonian also describes an array of
Josephson junctions where there is a phase shift of $\pi $ across a fraction
of the junctions as in models of d-wave ceramic superconductors \cite
{kawali,kusmartsev,dominguez}.

To study the vortex dynamics it is convenient to rewrite the above
Hamiltonian in the Coulomb-gas representation

\begin{equation}
H_{cg}=2\pi ^{2}J\sum_{r,r^{\prime }}(n_{r}-f_{r})G_{rr^{\prime }}^{\prime
}(n_{r^{\prime }}-f_{r^{\prime }})
\end{equation}
which can be obtained  following a standard procedure \cite{jose} 
in which Eq. (1) is replaced by a periodic Gaussian model separating spin-wave
and vortex variables. The Coulomb interaction is given by 
$G_{rr^{\prime }}^{\prime }=G(r-r^{\prime \;})-G(0)$, 
where $G$ is the lattice Green's function.

\begin{equation}
G(r)=\frac{1}{L^{2}}\sum_{k}\frac{\exp (ik\cdot r)}{4-2\cos k_{x}-2\cos k_{y}}
\end{equation}
and $L$ is the system size. $G^{\prime }(r)$ diverges as $2\pi \log |r|$ at
large separations $r$. The vortices are represented by integer charges $n_{r} $ 
at the sites $r$ of the dual lattice and the frustration effects of $A_{ij} $ 
by quenched random charges $f_{r}$ given by the directed sum of $A_{ij}$ around 
the plaquette, $f_{r}=\sum A_{ij}/2\pi $. The charges are
constrained by the neutrality condition $\sum_{r}(n_{r}-f_{r})=0$. For the
XY spin glass, the charges $f_{r}$ have a correlated random distribution of
integer and half integer values. Other random distributions can represent
different models. A uniform distribution describes the gauge glass model 
\cite{hyman,fisher91} where $A_{ij}$ in Eq. (1) is a continuous variable in
the range $[0,2\pi ]$ and an uncorrelated continuous distribution can
describe arrays of superconducting grains with random flux \cite{gk86} or
arrays of mesoscopic metallic grains with random offset charges \cite{eg98}.

We study the nonequlibrium response of the vortices in the XY spin glass by
Monte Carlo simulations of the Coulomb gas under an applied electric field 
\cite{hyman}. An electric field $E$ represents an applied force acting on
the vortices and gives an additional contribution to the energy in Eq. (2)
as $\sum_{r}E\ n_{r}\ x_{r}$ for $E$ in the $x$ direction. A finite $E$
sets an additional length scale \cite{hyman} in the problem since thermal 
fluctuations alone, of typical energy $kT$, leads to a characteristic 
length $l\sim kT/E$ over which single charge motion is possible. Thus, 
increasing $E$ will probe smaller length scales. Crossover effects are 
then expected when $l$ is of the order of the relevant correlation length for 
independent charge motion. As vortex motion leads to phase incoherence we thus 
expect that the scaling behavior of the nonlinear response will probe the spin glass 
correlation length $\xi _{s}$ of the original model and allow an estimate of the thermal
critical exponent $\nu _{s}$. This dynamical approach complements previous
equilibrium calculations in the vortex representation of the XY spin glass
where only the chiral  glass correlation length was studied \cite{by96}.

In the  dynamical simulations, the Monte Carlo time is identified as 
the real time $t$ and we take the  unit of time $dt=1$ corresponding to 
a complete Monte Carlo pass through the lattice.  A Monte Carlo step consists of adding a 
dipole of unit charges and unit length to a nearest neighbor charge pair $(n_{i},n_{j})$, 
using the Metropolis algorithm. Choosing a nearest-neighbor pair ${i,j}$ at random,
the step consists in changing $n_i \rightarrow n_i-1$ and $n_j \rightarrow n_j +1$,
corresponding to the motion of a charge by a unit length from $i$ to $j$. If the
change in energy is $\Delta U $, the move is accepted with probability 
min$\{1,\exp(-\Delta U/kT)\} $. The external electric field $E$ biases the added dipole, 
leading to a current $I$ as the net flow of charges in the direction of the electric
field if the charges are mobile. The current $I$ is calculated as 
\begin{equation}
I(t)=\frac{1}{L} \sum_i \Delta Q_i(t)
\end{equation}
after each Monte Carlo pass through the lattice, where $L$ is lattice size and 
$\Delta Q_i(t) =1$ if a charge at site $i$ moves one lattice spacing in the 
direction of the field $E$ at time $t$, $\Delta Q_i(t) =-1$ if it moves in
the opposite direction and $\Delta Q_i(t) =0$ otherwise. 
Periodic boundary conditions are used. Most calculations were performed for $L=32$ 
and compared to a smaller system of $L=16$ but size dependence was not significant 
in the temperature range studied. The current density $J$ is defined as $J=I/L$.
The linear response is given by the linear conductance 
$G_{L}=\lim_{E\rightarrow 0}J/E$ which can be obtained from the 
fluctuation-dissipation relation as

\begin{equation}
G_{L}=\frac{1}{2kT}\int dt<I(0)I(t)>
\end{equation}
without imposing the external field $E$. In the calculations, the integral is replaced
by a sum of successive Monte Carlo sweeps through the lattice with the unit of time
$dt=1$. We use typically $4\times 10^4$ Monte Carlo steps to compute averages and $20$ 
different realizations of disorder.

\begin{figure}[tbp]
\centering\epsfig{file=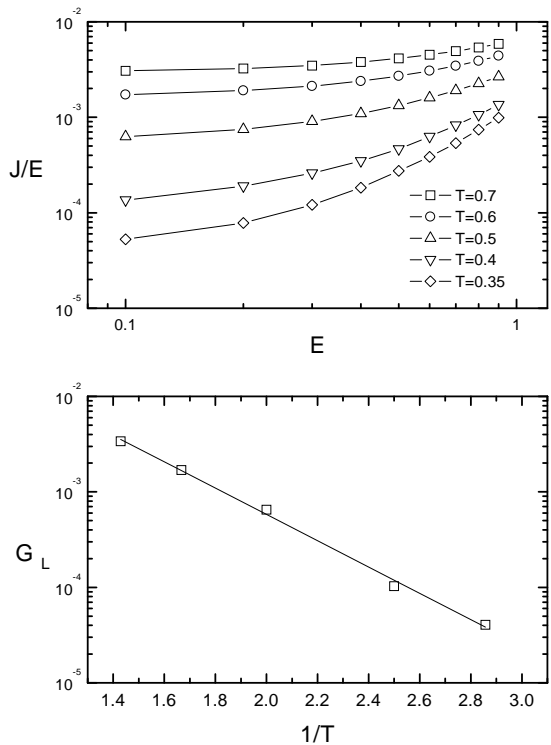,bbllx=5cm,bblly=0.9cm,bburx=16cm,
bbury=28cm,width=4.5cm}
\caption{(a) Nonlinear conductance $J/E$ as a function of temperature $T$.
(b) Arrhenius plot for the temperature dependence of the linear conductance 
$G_L$. }
\label{condcg}
\end{figure}

To analyze the numerical results we need a scaling theory of the nonlinear
response near a second-order phase transition. A detailed scaling theory 
has been  described in the context of  the current-voltage  characteristics 
of  vortex-glass models \cite{hyman} but it can be directly 
applied to the present case. Since the glass transition
occurs at $T=0$ with a power-law divergent correlation length $\xi
\propto T^{-\nu }$ and the external field introduces an additional length
scale $l \propto kT/E$, the dimensionless ratio  $E/J G_{L}$ 
can be cast into a simple scaling form \cite{hyman} in terms of the 
dimensionless argument $\xi/l$,
\begin{equation}
J/EG_{L}=F(E/T^{1+\nu })
\end{equation}
where $F$ is a scaling function with $F(0)=1$. This scaling form indicates
that a crossover from linear behavior, when $F(x) \sim 1 $, to nonlinear behavior,
when $F(x) >>1$, is expected to occur when $x \sim 1$ which 
leads to a characteristic field $E_{c}\propto T^{1+\nu }$ at which nonlinear
behavior sets in.

The nonlinear response $J/E$ and an Arrhenius plot of the linear conductance 
$G_{L}$ are shown in Fig. 1. The data shows the expected behavior for a $T=0$
transition. The ratio $J/E$ in Fig. 1(a) tends to a finite value for small $E$, 
corresponding to the linear conductance $G_{L}$ in Fig. 1(b) with an
activated behavior. This activated behavior is consistent with a zero temperature
transition and finite correlation length at nonzero temperatures which leads to 
a finite energy  barrier $U$ for vortex motion. In general, an energy barrier 
exponent \cite{fisher91} $\psi $ can also be defined from $U\sim \xi ^{\psi }$ 
for a temperature dependent energy barrier. The pure Arrhenius activated behavior 
in Fig. 1(b)  is consistent with an exponent $\psi \sim 0$. As can be seen from  Fig. 1(a),  
there is a smooth crossover from linear behavior, when $J/E$ is roughly a constant,
to nonlinear behavior for increasing $E$ at each temperature which  appears at 
smaller $E$ for  decreasing temperatures in agreement with the expected crossover
behavior at a characteristic field $E_{c}\propto T^{1+\nu }$ .

We now verify the scaling hypothesis of Eq. (6) and obtain a numerical estimate 
of the thermal correlation length exponent $\nu $ using two different methods. 
Fig. 2(a) shows the temperature dependence of $E_{c}$ defined as the value of $E$ 
where $E/JG_{L}$ starts to deviate from a fixed value of $2$. From the 
expected power-law  behavior for the crossover field $E_{c}\propto T^{1+\nu }$ 
we obtain a direct estimate of $\nu =1.4(2)$ in a log-log plot. From a scaling plot of 
the nonlinear response according to Eq. (6), $\nu$ can also be obtained
by adjusting its value so that the best data collapse is obtained as shown in Fig. 2(b).
The data collapse supports the scaling behavior and provides an independent estimate 
of $\nu =1.3$. From the two independent estimates we finally obtain $\nu =1.35\pm 0.2$. 

Our estimate of $\nu =1.35\pm 0.2$  from the scaling analysis of the 
vortex response is consistent with previous estimates of the 
spin glass correlation length exponent $\nu_s$ obtained in the phase representation 
of the XY spin glass \cite{kawamura,ray,jain}. These calculations give numerical 
estimates with comparable uncertainties ranging from $\nu_s=1$ to $1.2$. 
It also agrees with a recent calculation in the phase representation of Eq. (1), 
$\nu =1.1\pm 0.2$, using the same scaling analysis \cite{eg98}. This suggests
that the relevant length scale  for vortex dynamics is set by the spin glass
correlation length which determines short range phase coherence. Since the 
chiral glass correlation length exponent has been estimated to be significantly
larger, in the range \cite{kawamura,ray,by96,kosterlitz} $\nu_c=1.8 $ to $2.6$, 
it also supports the scenario in which the phase and chiral variables in the XY 
spin glass are decoupled on large length scales and order with different correlation
length exponents. However, as the errorbars of these estimates are
significant large, a single critical exponent \cite{hilhorst,kosterlitz} may
not be completely ruled out on pure numerical grounds and further work will 
be necessary to completely settle this issue. 

\begin{figure}[tbp]
\centering\epsfig{file=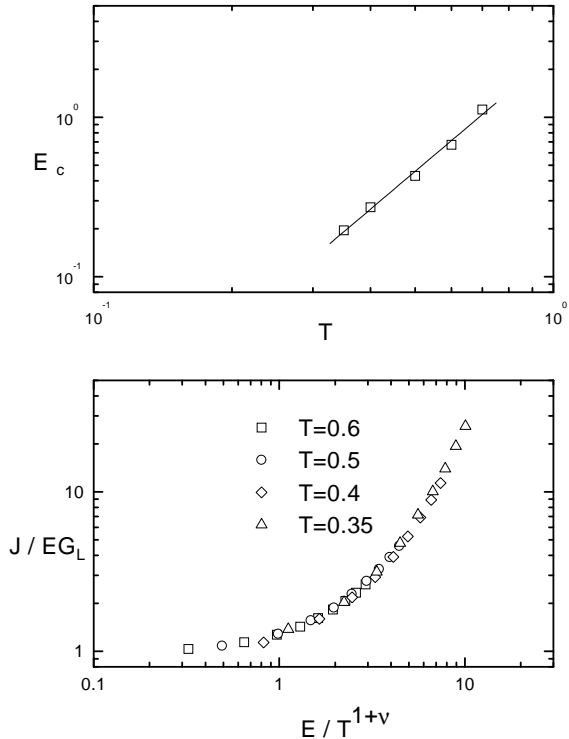,bbllx=5cm,bblly=0.9cm,bburx=16cm,
bbury=28cm,width=4.5cm}
\caption{(a) Crossover field $E_{c}$ as a function of temperature. (b)
Scaling plot $J/E G_L$ $\times$ $E/T^{1+\nu} $ for $\nu=1.3$.}
\label{scalcg}
\end{figure}

It should be noted that the decoupled scenario for spin and chiral variables near the  
same transition temperature, suggested by our numerical results, does not contradict 
general arguments of renormalization-group theory \cite{cardy} which allows for the
possibility of nontrivially decoupled fixed points. In fact, since the 
model has continuous and Ising-like symmetries, one would expect that the 
effective Hamiltonian describing the critical behavior could be  written in terms of 
a disordered ferromagnetic XY-spin model, with a zero-temperature transition,  coupled to 
an Ising spin glass model, representing  phase coherence and chiral degrees of freedom, 
respectively. Ferromagnetic XY spin models with a zero temperature transition do exist as, for example, 
in diluted XY models at percolation threshold \cite{stauffer,gd97}. Although the exact 
form of the effective XY and Ising Hamiltonians and the coupling  term are not known, 
the continuous and discrete symmetries  of the model are consistent with an energy 
density coupling  of  the form $\sum_r E_s(r) E_c(r)$, where $E_s$ and $E_c$ are the 
local energy densities for the XY spins  and chirality, respectively. Such a coupling term is 
known to occur in the effective Hamiltonian of frustrated XY models with 
weak disorder \cite{gk86}. For a stable decoupled fixed point, the coupling term should 
be an irrelevant perturbation, corresponding to an eigenvalue 
$\lambda=2-x < 0$  evaluated at the unperturbed fixed point, where $2x$ is the correlation 
function exponent.  Using the scaling relation \cite{cardy} $x=2-1/\nu$ for the energy 
density correlations, and the proposed numerical values for the correlation length 
exponents $\nu_s =1.3$ and $\nu_c=2$, we find indeed that $\lambda= 2-x_s-x_c < 0 $ as 
required for a stable decoupled fixed point. If the transition at $T=0$ corresponds
to a decoupled fixed point then phase and chiral variables can order with different
correlation length exponents. These arguments, by no means, show that a decoupled transition
is actually realized in the XY spin glass but it makes plausible the assumption of
distinct  divergent correlation lengths at the same transition temperature used in our
analysis of the numerical data. 
 
Finally, our calculation for the two-dimensional XY spin glass, which 
indicates  that vortex dynamics probe mainly the spin glass correlation length rather 
than the chiral glass correlation length, also suggests  that the finite temperature resistive 
transition observed recently by Wengel and Young \cite{wengel}
in numerical simulations in the vortex representation of the tree-dimensional XY spin glass
model should be attributed to spin glass ordering. This is in fact 
consistent with more recent  calculation \cite{grempel} indicating that the 
lower critical dimension for spin-glass  ordering may be just above $3$.

\bigskip

This work was supported by Funda\c{c}\~{a}o de Amparo \`{a} Pesquisa do
Estado de S\~{a}o Paulo, FAPESP (Proc. 99/02532-0).

\end{document}